\documentclass[aps,pra,showpacs, two column]{revtex4}

\begin{document}

\title{Generation of entangled tripartite states in three identical cavities}

\author{Moorad Alexanian}
\email[]{alexanian@uncw.edu}

\affiliation{Department of Physics and Physical Oceanography\\
University of North Carolina Wilmington\\ Wilmington, NC
28403-5606\\}

\date{\today}

\begin{abstract}
The generation of entanglement between three identical coupled cavities, each containing a single three-level atom, is studied when the cavities exchange two coherent photons and are in the $N$=2, 4, and 6 manifolds, where $N$ represents the maximum number of photons possible in any one cavity.  The combined states of the atom and the photon in a cavity is given by a qutrit for $N=2$, a five-dimensional qudit for $N=4$, and a seven-dimensional qudit for $N=6$. The conservation of the operator  $\hat{N}$ for the interacting three-cavity system limits the total number of tripartite states to only 6, 18, and 38, rather than the usual $3^3=27$, $5^3=125$, and $7^3 =343$ states for $N$=2, 4, and 6, respectively. The deterministic generation of entanglement from general initially unentangled tripartite states is studied in the limit of large hopping strength, where all the solutions are analytic and given in terms of exponential functions. Several types of resulting tripartite entanglement are analyzed in order to obtain maximally entangled states.
\end{abstract}


\maketitle {}

\section{Introduction}

Quantum entanglement is essential for quantum information communication and processing protocols in quantum cryptography \cite{E91}, dense coding \cite{BW92}, teleportation \cite{BBC93}, and entanglement swapping \cite{PBW98}, which can be used to realize quantum repeaters \cite{BDC98}. Entanglement can be achieved via two interacting quantum systems \cite{KMW95} or by an appropriate joint measurement of two systems \cite{W94}. An important use of entanglement is the sharing of multipartite states distributed amongst several different parties that are separated by large distances where the performance of local measurements on their respective subsystems results in the global broadcasting of the outcomes of local measurements. The latter is used in standard quantum key distribution \cite{BB84} and quantum secret sharing \cite{HBB99}. Accordingly, generating multipartite entanglement is an important objective in experimental quantum systems. For instance, experimental realizations of four-photon entanglement and high-fidelity teleportation \cite{PDG01}, entangled states of two and four trapped ions \cite{SKK00}, the entanglement of two Rydberg atoms in microwave cavity quantum electrodynamics (QED) \cite{HMN01}, optically  induced entanglement  of excitons in a single quantum dot \cite{CBS00}, creation of Greenberger-Horne-Zeilinger (GHZ) states with up to 14 qubits \cite{MSB11}, and the entanglement of a six-photon symmetric Dicke state \cite{WKK09}. On the theoretical side, many methods based on cavity QED have been proposed, for instance, a hybrid quantum correlated tripartite system formed by an optical cavity and a microwave cavity \cite{BVT11}, generating $n$-qubit GHZ entangled states with a three-level qubit system and $(n - 1)$ four-level qubit systems in a cavity \cite{Y11}, and the generation of maximally entangled GHZ state as a ground state of a three spin system \cite{SH11}. In general, the more particles or states can be entangled, the more clearly non-classical effects are exhibited and the more useful such entangled states are for quantum applications \cite{PBD00, M90}.

In the study of three-level atoms, a transformation was introduced \cite{AB95} whereby the three-level atom was reduced to a corresponding two-level atom of the Jaynes-Cummings type albeit with two-photon rather than single-photon transitions. This model has been used in cavity QED to generate ``macroscopic" qubits \cite{AB02} and in the scattering of two coherent photons inside a one-dimensional coupled-resonator waveguide that operates as an ideal quantum switch \cite{MA10}.   The model has been applied to two-photon exchange between two separate cavities \cite{MA11} with each cavity containing a three-level atom in a cascade (or ladder) configuration \cite{ABC98} and coupled via a two-photon hopping interaction \cite{MA11}. The coupling for two-photon hopping would require a nonlinear media that may be governed by a quantum Kerr-type interaction \cite{DELL06}. The latter work \cite{MA11} was restricted to the $N=2$ manifold, where $N$ denotes the maximum number of photons possible in a given cavity. More recently, we extended our study to the dynamics of the $N=4$ manifold and showed how the temporal development of the coupled two-cavity system generates maximally entangled states in both the $N=2$ and $N=4$ manifolds from an initially unentangled state \cite{MA84}.

In order to realize maximally entangled states, one must consider high Q-factor cavities and excited states that are long-lived. Owing to angular momentum and parity conservation, the simultaneous emissions of two photons possess increased lifetimes and thus are essentially metastable states. For instance, single-photon atomic transitions are of the order of $10^{-8}$s whereas the two-photon emissions have much longer lifetimes, of the order of fractions of a second. This makes the study of two-photon processes quite interesting as compared to single-photon transitions. In addition, two-photon spectroscopy has been recently used to probe the hybridization between a superconducting phase qubit and an intrinsic two-level system coupling to the qubit circuit \cite{BML10}. The detailed data on experiments allow the mapping out of this hybrid system, combining two coherent quantum systems that are fundamentally different in nature \cite{BML10}. The experimental realization of the transmon qubit, which is an improved superconducting charge qubit derived from the Cooper pair box, considers two-photon transitions that could help design more robust quantum computers \cite{SHK08}. A recent study of the possibility of coherent reversal information transfer between superconducting charge qubits and mesoscopic ultracold atomic ensembles coupled to a microwave coplanar waveguide is based on two-photon optically excited Rydberg transitions--from the ground state to the Rydberg state via a nonresonant intermediate Rydberg state \cite{PBK09}. The experimental implementation of the latter proposal is currently underway to demonstrate hybridization of solid-state and atomic quantum devices \cite{PBK09}.

In this paper, we consider a system of three identical cavities each with a three-level atom where the cavities are coupled to each other by a hopping interaction that exchanges two coherent photons. We consider the generation of entangled tripartite states from initially unentangled states in the manifolds $N$=2, 4, and 6, which lie in a 6-, 18-, and 38-dimensional Hilbert space, respectively. Our system of three identical cavities consists of three atoms and one, two, and three pairs of photons in the manifolds $N$=2, 4, and 6, respectively. Accordingly, generalized measurements \cite{JC02}, by using only linear elements and particle detectors, may be possible on the generated entangled pure states.

\section{Hamiltonian}

The Hamiltonian for the three coupled cavities \cite{MA84} is given by
\[
H= \sum_{i=1}^{3} \big{[}H^{(i)} - H_{0}^{(i)}\big{]} + \hbar \xi (a_{1}^{\dag 2} a_{2}^2 + a_{2}^{\dag 2} a_{1}^2)
\]
\begin{equation}
+ \hbar \xi (a_{1}^{\dag 2} a_{3}^2 + a_{3}^{\dag 2} a_{1}^2) + + \hbar \xi (a_{2}^{\dag 2} a_{3}^2 + a_{3}^{\dag 2} a_{2}^2),
\end{equation}
where the first term on the right-hand-side represents the sum of the Hamiltonian for each cavity and the succeeding terms are the hopping interactions coupling the three cavities in a symmetric fashion. Therefore, the Hamiltonian (1) is symmetric in the three identical cavities.

The eigenvalues $E^{\pm}_{n}$ and eigenfunctions
$|\Psi^{\pm}_{n}\rangle^{(i)}$ of $H^{(i)}$ are best given in terms of the
dressed-atom representation. One has \cite{ABC98} that
\begin{eqnarray}
|\Psi^{+}_{n}\rangle^{(i)} = \textup{sin} \theta_{n}|e,n\rangle^{(i)} +
\textup{cos}\theta_{n}|g,n+2\rangle^{(i)} \nonumber\\|\Psi^{-}_{n}\rangle^{(i)} =
\textup{cos}\theta_{n}|e,n\rangle^{(i)} - \textup{sin}\theta_{n}|g,n+2\rangle^{(i)},
\end{eqnarray} where $|g\rangle^{(i)}$ and $|e\rangle^{(i)}$ are the ground and the excited atomic
states, respectively, $|n\rangle^{(i)}$ is the photon number eigenstate and
\begin{eqnarray}
\textup{cos}\theta_{n}=
\frac{r(n+2)^{1/2}}{[n(r^2+1)+2r^2+1]^{1/2}}\nonumber\\
\textup{sin}\theta_{n}=\frac{(n+1)^{1/2}}{[n(r^2+1)+2r^2+1]^{1/2}},
\end{eqnarray}
with $r\equiv g_{1}/g_{2}$.

The respective eigenvalues are given by
\[
E^+ _{n} =  \hbar \omega(n+1) + \frac{E_{g} + E_{e}}{2} -\frac{\Delta}{2}
\]
\begin{equation}
+\frac{1}{2}\{\Delta^2+ 4\hbar^2[g_{1}^2(n+2) +g_{2}^2(n+1)]\}^{1/2}
\end{equation}
\[
E^-_{n} = \hbar\omega(n+1)+\frac{E_{g}+ E_{e}}{2},
\]
where $g_{j}$ ($j=1,2$) are the atom-photon coupling constants in the three-level atom, $E_{g}$ is
the energy of the ground state, $E_{e}$ is the energy of the excited
state, and $\Delta=(E_{g}-E_{2})+
\hbar\omega=(E_{e}-E_{2})-\hbar\omega$ is the detuning parameter
of the mid level with energy $E_{2}$ of the three-level atom in a cascade (or ladder) configuration. The eigenenergies of the dressed states depend on $n$ reflecting the exact treatment of intensity-dependent dynamic Stark shifts.

The eigenstates $|\Psi^{\pm}_{n}\rangle^{(i)}$ are simultaneous eigenstates of $H^{(i)}$ with eigenvalues $E^{\pm}_{n}$, of $H_{0}^{(i)}= \hbar\omega(a_{i}^\dag a_{i}+\sigma_{ee}^{(i)}-\sigma_{gg}^{(i)}) + (E_{g}+ E_{e})/2$ with eigenvalues $E^-_{n}$, and of $\hat{N}^{(i)}=a_{i}^\dag
a_{i}+\sigma_{ee}^{(i)}-\sigma_{gg}^{(i)}+1$ with eigenvalues $n+2$. The operator $\hat{N} = \sum_{i=1}^{3} \hat{N}^{(i)}$, associated with the three interacting cavities, is a constant of the motion and in this paper we consider the manifolds with eigenvalues $N=2$, 4, and 6. The set of eigenstates $|\Psi^{\pm}_{n}\rangle^{(i)}$, with $n= 0, 1, 2\cdots,$ together with the states $|g,0\rangle^{(i)} = -|\Psi^{-}_{-2}\rangle^{(i)}$ and $|g,1\rangle^{(i)}=|\Psi^{+}_{-1}\rangle^{(i)}$, where the first index refers to the
ground state and the second to the photon-number occupation, forms a complete basis.

\section{N=2 manifold}

The Hilbert space  of the $N=2$ manifold corresponds to each atom in the three-cavity system being described by the qutrit $|g,2\rangle$, $|g,0\rangle$, and $|e,0\rangle$. However, owing to the constancy of the total operator $\hat{N}=\hat{N}^{(1)}+\hat{N}^{(2)}+\hat{N}^{(3)}$, the space is spanned by only 6 rather than $3^3= 27$ vectors.

The most general unentangled initial tripartite state in the $N=2$ manifold is
\begin{equation}
|\psi(0)\rangle = |g,0\rangle^{(1)} |g,0\rangle^{(2)} \Big{[} a |g,2\rangle^{(3)} + b |e,0\rangle^{(3)}\Big{]},
\end{equation}
where $|a|^2 + |b|^2=1$ and $|g\rangle^{(i)}$ and $|e\rangle^{(i)}$ are the ground and the excited atomic states, respectively, and $|n\rangle^{(i)}$ is the photon number eigenstate with $i =1, 2, 3$ denoting the cavity. The initial state (5) is only symmetric in the exchange $1 \leftrightarrow 2$ and cavity 3 is in an entangled state given by the linear superposition of dressed states $|\Psi^{\pm}_{0}\rangle^{(3)}$. Owing to the symmetry of the Hamiltonian (1), the general solution preserves that symmetry and is given by
\[
|\psi(t)\rangle =  |g,0\rangle^{(1)} |g,0\rangle^{(2)} \Big{[} A(t) |g,2\rangle^{(3)} +  C(t)   |e,0\rangle^{(3)}\Big{]}
\]
\begin{equation}
+ \frac{1}{\sqrt{2}} B(t) \Big{[} |g,0\rangle^{(1)} |g,2\rangle^{(2)} + |g,2\rangle^{(1)} |g,0\rangle^{(2)}  \Big{]} |g,0\rangle^{(3)}.
\end{equation}
The Schr\"{o}dinger equation of the motion $i\hbar \frac{\partial}{\partial t} |\psi\rangle = H |\psi\rangle$ gives, with the aid of Eqs. (1) and (6),
\[
i\dot{A} =  A + 2 \sqrt{2}  \xi B + \tan\theta_{0} C
\]
\begin{equation}
i\dot{B} = 2 \sqrt{2} \xi A + (1+ 2\xi) B,
\end{equation}
\[
i \dot{C} = \tan \theta_{0} A + \tan^2 \theta_{0} C,
\]
with initial conditions $A(0)=a$, $B(0)=0$, and $C(0)=b$. In Eq. (7), we have introduced the dimensionless time $[(E_{0}^{+} - E_{0}^{-}) \cos^2 \theta_{0}] t/\hbar \rightarrow t $ and the dimensionless hopping coupling $ \hbar\xi/[ (E_{0}^{+} - E_{0}^{-}) \cos^2 \theta_{0} ] \rightarrow \xi$ with the angle $\theta_{0}$ representing the mixing angle (3) for the atomic dressed states. The system of equations (7) gives rise to analytic solutions for arbitrary values of the parameters $r$ and $\xi$. However, we shall consider the case of large hopping strength $\xi\gg 1$ when the exchange of two coherent photons occurs at a much faster rate than the rate of atomic transitions. In that case, one has the separate probability conservation
\begin{equation}
|A(t)|^2 + |B(t)|^2 \approx |a|^2  \hspace{0.2in} \textup{and} \hspace{0.2in}  |C(t)|^2 \approx |b|^2 .
\end{equation}
Note that the amplitude $C(t)$ associated with cavity 3 being in the excited state is separately conserved according to (8). This follows directly from considering the limit of large hopping strength, viz, $\xi\gg 1$. Similarly, for the amplitudes $A(t)$ and $B(t)$ associated with all three cavities being in the ground state that are separately conserved as indicated in (8).

Therefore, one achieves maximal entanglement for times $T$ when $|A(T)|^2 + |C(T)|^2$ is a minimum, that is, $|A(T)|$ is a minimum with $C(0)= b=0$. The solution for the system of equations (7) is then for $\xi\gg 1$
\[
A(t)  \approx \frac{1}{3} \hspace{0.02 in}    (e^{ -4 i\xi t}+  2 e^{2 i  \xi t}),
\]
\begin{equation}
B(t) \approx  \frac{\sqrt{2} }{3} \hspace{0.02 in}  (e^{- 4 i\xi t} - e^{2 i  \xi t}),
\end{equation}
\[
C(t) = 0,
\]
where the amplitudes $A(t)$ and $B(t)$ are periodic with period $T=\pi/\xi$. Result (9) agrees with the general results given by (A3) in Appendix A for the corresponding amplitudes. Note that the state (6) becomes unentangled for times $\tau$ such that $\xi \tau = \pi n/3$, $n=1, 2, \cdots$, that is, $B(\tau)=0$. Our characterization of maximal entanglement for the $N=2$ manifold given above is quite consistent with the notion of geometric measure of entanglement \cite{S95, M11, WG03},
\begin{equation}
E(|\psi(t)\rangle) = \min_{|\phi\rangle  \in \hspace{0.02in}\textup{PROD}} -\log_{2} (|\langle\phi|\psi(t)\rangle|^2),
\end{equation}
where PROD is the set of product states in the $N=2$ manifold. Now
\begin{equation}
\max_{|\phi\rangle  \in \hspace{0.02in}\textup{PROD}}|\langle\phi|\psi(t)\rangle|^2 = |A(t)|^2 + |C(t)|^2
\end{equation}
\[
= \frac{1}{9} \Big{[}5 + 4 \cos(6\xi t)\Big{]} |a|^2 + |b|^2.
\]
Therefore, the measure of entanglement  $ 0 \leq E(|\psi(t)\rangle) \leq  \log_{2}9 \approx 3.170$, where the lower bound corresponds to times $ \xi \tau =\pi l/3$, $l=0, 1, 2, \cdots$, and the upper bound to times $ \xi \mu = (2 n +1) \pi /6$ with $n= 0, 1, 2, \cdots$ and $|a|=1$. Note that the state (6) has an entangled two-qubit subsystem and, hence, does not admit Schmidt decomposition.

The average dwell or sojourn time the system spends in the entangled state governed by the amplitude $B(t)$ is given by $\frac{1}{\pi} \int_{0}^{\pi} |B(t)|^2 dt = \frac{4}{9} |a|^2$.
Note that the initial state (5) possesses less symmetry than the Hamiltonian (1). Actually, there are only two symmetric states, viz., $\frac{1}{\sqrt{3}}\sum_{P}  P |e,0\rangle^{(1)}|g,0\rangle^{(2)}|g,0\rangle^{(3)}$ and $\frac{1}{\sqrt{3}}\sum_{P}  P |g,2\rangle^{(1)}|g,0\rangle^{(2)}|g,0\rangle^{(3)}$, where the sums are over even permutation, and no antisymmetric states are possible. Both symmetric states are entangled since there are no unentangled, symmetric state in the $N=2$ manifold. Therefore, the six-dimensional space for the $N=2$ manifold is spanned, in addition to these two symmetric states, by four asymmetric vectors.

\section{N=4 manifold}

The Hilbert space  of the $N=4$ manifold corresponds to each atom in the three-cavity system being described by the five-dimensional qudit $|g,4\rangle$, $|g,2\rangle$, $|g,0\rangle$, $|e,0\rangle$, and $|e,2\rangle$. However, the space is spanned by only 18 vectors rather than $3^5= 125$ vectors owing to the constancy of the operator $\hat{N}$ and so a general state is given by (B1) in Appendix B. This manifold is spanned by 5 symmetric, one antisymmetric, and 12 asymmetric states, which treat all three cavities on the same footing. The most general unentangled  initial states in the $N=4$ manifold are
\begin{equation}
|\psi(0)\rangle = |g,0\rangle^{(1)}  |g,0\rangle^{(2)}  \Bigg{[} a |g,4\rangle^{(3)}  + b|e,2\rangle^{(3)}  \Bigg{]},
\end{equation}
where $|a|^2 + |b|^2=1$, (12) is symmetric under the exchange $1 \leftrightarrow 2$, cavity 3 is in an entangled state, and
\begin{equation}
|\psi(0)\rangle = |g,0\rangle^{(1)}    \Bigg{[} a |g,2\rangle^{(2)}  + b|e,0\rangle^{(2)}  \Bigg{]} \Bigg{[} c |g,2\rangle^{(3)}  + d |e,0\rangle^{(3)}  \Bigg{]},
\end{equation}
where $|a|^2 + |b|^2=1$, $|c|^2 + |d|^2=1$, and both cavities 2 and 3 are in entangled states. For $c=a$ and $d=b$, the initial state(13) is symmetric under the interchange $2\leftrightarrow 3$.

\subsection{Initial state Eq. (12)}

The initial state (12) is symmetric under the exchange of cavities $1\leftrightarrow 2$ and so the time dependent state $|\psi(t)\rangle$  lies in an 11-dimensional subspace rather than the possible 18 since $A(t)=G(t)$, $B(t)=P(t)$, $D(t)=S(t)$, $E(t)=N(t)$, $W(t)=J(t)$, $L(t)=U(t)$, and $M(t)=R(t)$ in (B1). On the other hand, the initial state (13) for $ad\neq bc \neq 0$  is not symmetric in the exchange $2\leftrightarrow 3$  and so lies in the full 18-dimensional space.   However, for  $ad = bc$ the initial state (13) is symmetric in the exchange $2\leftrightarrow 3$ and so the initial state lies also in an 11-dimensional subspace. Owing to the complexity of the full 18-dimensional vector space, we shall consider the limit of large hopping strength $\hbar \xi/[(E^{+}_{2} - E^{-}_{2}) \cos^2 \theta_{2}] \gg 1$ when the exchange of two coherent photons between the cavities is much faster than atomic transitions.

Consider first the solution of the Schr\"{o}dinger equations (B2)--(B7) for the initial unentangled state (12), that is, $F(0) =a$ and $K(0)=b$ in (B1). The general 18-dimensional vector (B1) in the $N=4$ manifold must be symmetric in the interchange of cavities $1\leftrightarrow 2$ and so the only nonzero amplitudes are given by $G(t)=A(t)$, $P(t)=B(t)$, $N(t)=E(t)$, $C(t)$,  $F(t)$, and  $K(t)$. Therefore, the solution is
\[
|\psi(t)\rangle \approx  A(t) \Bigg{[}|g,4\rangle^{(1)} |g,0\rangle^{(2)} + |g,0\rangle^{(1)} |g,4\rangle^{(2)} \Bigg{]} |g,0\rangle^{(3)}
\]
\[
+\Bigg{[}|g,0\rangle^{(1)} |g,2\rangle^{(2)} + |g,2\rangle^{(1)} |g,0\rangle^{(2)} \Bigg{]}\times
\]
\begin{equation}
 \times\Bigg{[} B(t)  |g,2\rangle^{(3)} +E(t) |e,0\rangle^{(3)}\Bigg{]}
\end{equation}
 \[
+ C(t) |g,2\rangle^{(1)}|g,2\rangle^{(2)}|g,0\rangle^{(3)}
\]
\[
+ |g,0\rangle^{(1)}|g,0\rangle^{(2)} \Bigg{[} F(t) |g,4\rangle^{(3)} + K(t)|e,2\rangle^{(3)} \Bigg{]},
\]
where
\[
A(t) = \frac{1}{15}a \Bigg{[}  3 e^{8 i \xi t} -2 e^{6i \xi t} + 2 e^{-12 i \xi t} - 3 e^{ -4 i \xi t}  \Bigg{]}
\]
\[
B(t) = \frac{\sqrt{6}}{15} a \Bigg{[} - e^{8i \xi t} -  e^{6 i \xi t} +  e^{ -12 i \xi t} +  e^{-4 i \xi t} \Bigg{]}
\]
\begin{equation}
C(t) = \frac{\sqrt{6}}{15} a \Bigg{[} - e^{8i \xi t} +2  e^{6 i \xi t} +  e^{ -12 i \xi t} -2  e^{-4 i \xi t} \Bigg{]}
\end{equation}
\[
E(t) = \frac{1}{3} b \Bigg{[}   -  e^{ 2 i \xi t} + e^{-4 i \xi t}\Bigg{]},
\]
\[
F(t) = \frac{1}{15} a\Bigg{[} 3 e^{8i \xi t} +4  e^{6 i \xi t} + 2 e^{ -12 i \xi t} +6  e^{-4 i \xi t} \Bigg{]}
\]
\[
K(t) = \frac{1}{3} b \Bigg{[} 2  e^{ 2 i \xi t} + e^{-4 i \xi t} \Bigg{]}.
\]
The amplitudes in (15) are periodic with period $T= \pi/\xi$. Note that for any probability amplitude $X(t)$, if $\frac{d^{n} X(t)}{dt^n}|_{t=0} =0$ for $n =0, 1, 2, \cdots$, then $X(t) \equiv 0$ since the solutions of the dynamical equations are entire functions of $t$. The latter has been used in Eqs (B2)--(B7) to arrive at the five nonzero amplitudes given in (15) for solution (14). Note that $2|E(t)|^2 +|K(t)|^2 =|b|^2$ and $2|A(t)|^2 +2|B(t)|^2 +|C(t)|^2 + |F(t)|^2 =|a|^2$ are separately conserved with the sum giving the overall probability of unity. This feature of disjoint sectors of the Hilbert space, determined by the number of atoms in the excited state, is a direct result of large hopping strength and occurs in all the different manifolds.

One obtains maximal entanglement for the initial state (12) with $a=0$ by maximizing $|E(t)|^2$, or what is the same by minimizing $|K(t)|^2$, which occurs at a time $\tau $ given by $\tau \xi = \pi/6$, where $|K(\tau)|^2 = 1/9$ and $2|E(\tau)|^2 = 8/9$ . The measure of entanglement is $\log_{2}9 \approx 3.170$ that is the same as for the $N=2$ manifold albeit for different states.

On the other hand, for the initial state with $b=0$ in Eq. (12), that is, all four photons are initially in cavity 3, there are several interesting entangled states obtained, for instance, by either maximizing or minimizing $|A(t)|^2 + |B(t)|^2$, which is the same as minimizing or maximizing $|C(t)|^2 + |F(t)|^2$. For instance, minimization of $|C(t)|^2 + |F(t)|^2$ occurs at $\xi \tau \approx 0.2094, 0.8378$ with  $|C(\tau)|^2 + |F(\tau)|^2 \approx 0.1960$ resulting in a lower measure of entanglement $\log_{2}(1/0.1960) \approx 2.351$. The probabilities for the entangled states are  $2|A(\tau)|^2 \approx  .2251$ and $2|B(\tau)|^2 \approx  .5789$. Is is interesting that at $\xi \tau =\pi/3$ one has that $A(\tau) = B(\tau)=0$ with $C(\pi/3) =3\sqrt{2}(\sqrt{3}-i)/10$ and $F(\pi/3)= (1+ 3\sqrt{3}i)/10$. Therefore, the probability that all four photons remain in cavity 3 is $28\%$ with the remaining $72\%$ with cavities 1 and 2 sharing two photons equally.

\subsection{Initial state Eq. (13)}

Consider next the initial unentangled state (13), that is, $L(0) = b d$, $M(0) = bc$, $N(0)= a d$, and $P(0) = a c$, with the time evolution of the probability amplitudes given by Eqs. (B2)--(B7). Now $B(t)=C(t)$ and $G(t)=F(t)$ according to Eq. (B3), $D(t)= J(t)$ follows from Eq. (B4), Eq.(B5) gives that $E(t)=K(t)$, while $T(t) = U(t)=0$ from Eq. (B6), and Eq. (B7) implies that $R(t)=S(t)=W(t)=0$. The state of the system is then, according to (B1), given by
\[
|\psi(t)\rangle = A(t) |g,4\rangle^{(1)} |g,0\rangle^{(2)}|g,0\rangle^{(3)}
\]
\[
+ B(t)|g,2\rangle^{(1)}   \bigg{[}|g,0\rangle^{(2)}|g,2\rangle^{(3)}  +   |g,2\rangle^{(2)} |g,0\rangle^{(3)} \bigg{]}
\]
\[
 +D(t)\bigg{[} |g,2\rangle^{(1)}  |e,0\rangle^{(2)} +  |g,0\rangle^{(1)}  |e,2\rangle^{(2)}\bigg{]}|g,0\rangle^{(3)}
\]
 \[
 + E(t) \bigg{[} |g,2\rangle^{(1)}  |g,0\rangle^{(2)}|e,0\rangle^{(3)} +  |g,0\rangle^{(1)} |g,0\rangle^{(2)}|e,2\rangle^{(3)} \bigg{]}
\]
\begin{equation}
 + F(t) |g,0\rangle^{(1)} \bigg{[} |g,0\rangle^{(2)}|g,4\rangle^{(3)} + |g,4\rangle^{(2)}|g,0\rangle^{(3)}\bigg{]}
\end{equation}
\[
+ |g,0\rangle^{(1)}  \bigg{[} L(t) |e,0\rangle^{(2)}|e,0\rangle^{(3)}+M(t) |e,0\rangle^{(2)}|g,2\rangle^{(3)}
\]
\[
+ N(t) |g,2\rangle^{(2)}|e,0\rangle^{(3)} + P(t) |g,2\rangle^{(2)}|g,2\rangle^{(3)}\bigg{]},
\]
where
\[
A(t)= \frac{\sqrt{6}}{15} a c \Big{[} -e^{8 i \xi t} + 2 e^{6 i \xi t} -2 e^{- 4 i \xi t} + e^{- 12 i \xi t} \Big{]},
\]
\[
B(t) = \frac{1}{15} a c \Big{[} 2 e^{8 i \xi t} - 3 e^{6 i \xi t} -2 e^{- 4 i \xi t} + 3 e^{- 12 i \xi t} \Big{]},
\]
\[
D(t) =\frac{1}{3} b c \Big{[} - e^{2 i \xi t} + e^{- 4 i \xi t}  \Big{]},
\]
\[
E(t) = \frac{1}{3} a d \Big{[} - e^{2 i \xi t} + e^{- 4 i \xi t}  \Big{]},
\]
\begin{equation}
F(t)= \frac{\sqrt{6}}{15} a c \Big{[} -e^{8 i \xi t} - e^{6 i \xi t} + e^{- 4 i \xi t} + e^{- 12 i \xi t} \Big{]},
\end{equation}
\[
L(t) = b d,
\]
\[
M(t) = \frac{1}{3} b c \Big{[} 2 e^{2 i \xi t} + e^{- 4 i \xi t}  \Big{]},
\]
\[
N(t) = \frac{1}{3} a d \Big{[} 2 e^{2 i \xi t} + e^{- 4 i \xi t}  \Big{]},
\]
\[
P(t) =  \frac{1}{15} a c \Big{[} 2 e^{8 i \xi t} + 6 e^{6 i \xi t} + 4 e^{- 4 i \xi t} + 3 e^{- 12 i \xi t} \Big{]}.
\]
The solutions in (17) are periodic with period $T=\pi/\xi$. Note that $ |A(t)|^2 + 2|B(t)|^2 + 2|F(t)|^2 + |P(t)|^2=|a|^2|c|^2$, $|L(t)|^2 = |b|^2|d|^2$, $2|D(t)|^2  +|M(t)|^2 = |b|^2 |c|^2$, and $2|E(t)|^2  +|N(t)|^2 = |a|^2|d|^2$  are separately conserved with the sum giving the overall probability of unity. This feature of disjoint sectors of the Hilbert space for states with one, two, or none of the atoms in the cavities in the excited state is a direct result of considering the limit of large hopping strength.

Consider the case with symmetry under the cavity-interchange $2\leftrightarrow 3$, that is, $c=a$ and $d=b$ in Eq. (13), and so (17) gives that $D(t)=E(t)$ and $M(t)=N(t)$. For the initial state with $a=0$, the three-cavity system remains in its initial state, viz., $ |g,0\rangle^{(1)}|e,0\rangle^{(2)}|e,0\rangle^{(3)}$, since  for $\xi \gg 1$ the hopping interaction does not give rise to any temporal change. However, for the initial state $|g,0\rangle^{(1)}|g,2\rangle^{(2)}|g,2\rangle^{(3)}$, that is, $a=c=1$ and $b= d = 0$ in (13), one has
\[
|\psi(t)\rangle = A(t) |g,4\rangle^{(1)} |g,0\rangle^{(2)}|g,0\rangle^{(3)}
\]
\begin{equation}
+ B(t)|g,2\rangle^{(1)} \Bigg{[} |g,0\rangle^{(2)}|g,2\rangle^{(3)}  +   |g,2\rangle^{(2)} |g,0\rangle^{(3)} \Bigg{]}
\end{equation}
\[
 + F(t) |g,0\rangle^{(1)} \Bigg{[} |g,0\rangle^{(2)}|g,4\rangle^{(3)} + |g,4\rangle^{(2)}|g,0\rangle^{(3)}\Bigg{]}
\]
\[
+P(t)|g,0\rangle^{(1)} |g,2\rangle^{(2)}|g,2\rangle^{(3)},
\]
where the amplitudes are given in (17) with $a= c=1$ and $b= d=0$ and the normalization by  $ |A(t)|^2 + 2|B(t)|^2 + 2|F(t)|^2 + |P(t)|^2=1$. One obtains maximal entanglement for the state (18) if one minimizes $|A(t)|^2 + |P(t)|^2$. The same minimum occurs for $\xi \tau \approx  .1930, 0.8542, 1.2402$. One obtains the state where the respective probabilities are $|A(\tau)|^2 \approx 0.1070$, $2|B(\tau)|^2 \approx 0.2112$, $2|F(\tau)|^2 \approx 0.6060$, $|P(\tau)|^2 \approx 0.0759$. The measurement of entanglement is $-\log_{2}(|A(\tau)|^2 + |P(\tau)|^2) \approx 2.450$.

Is is interesting that at $\xi \tau =\pi/3$ one has that $B(\tau) = F(\tau)=0$ with $A(\pi/3) =\sqrt{6}(3-\sqrt{3}i)/10$ and $P(\pi/3)= (2+ \sqrt{3} i)/5$. Therefore, the probability that the system is in its initial state with two photons each in cavities 2 and 3 is $28\%$ while the probability of four photons in cavity 1 is $72\%$.

\section{N=6 manifold}

The seven-dimensional qudit associated with the states of each cavity in the $N=6$ manifold are $|g,6\rangle$, $|g,4\rangle$, $|g,2\rangle$, $|g,0\rangle$, $|e,0\rangle$,  $|e,2\rangle$, and $|e,4\rangle$, which gives rise to $7^3=343$ states for the three-cavity system. However, owing to the operator $\hat{N}$ being a constant of the motion, the relevant Hilbert subspace for the $N=6$ manifold is only 38-dimensional. The dimensionality is given by $1\times 10 + 3 \times 6 + 3\times3 +1\times 1 =38$, where the terms in the sum correspond to sectors $a$, $b$, $c$, and $d$, respectively (See Appendix C).  In sector $a$, all three cavities are in the ground state and there are six photons to be shared pairwise amongst the three cavities. In sector $b$, one of the cavities is in the excited state, two are in the ground state, and there are four photons to be shared pairwise. In sector $c$, two of the cavities are in the excited state, only one of the cavities is in the ground state, and there are two photons to be shared. Finally, in sector $d$, all three cavities are in the excited state and there are no photons to be shared amongst the three cavities.

The most general unentangled, initial states in the $N=6$ manifold are
\begin{equation}
|\psi(0)\rangle = \bigg{[} a|g,6\rangle^{(1)} + b|e,4\rangle^{(1)}\bigg{]} |g,0\rangle^{(2)} |g,0\rangle^{(3)},
\end{equation}
where $|a|^2 + |b|^2 =1$,
\begin{equation}
|\psi(0)\rangle = \bigg{[} a|g,4\rangle^{(1)} + b|e,2\rangle^{(1)}\bigg{]}  \bigg{[} c |g,2\rangle^{(2)} + d  |e,0\rangle^{(2)}\bigg{]} |g,0\rangle^{(3)},
\end{equation}
where $|a|^2 + |b|^2 =1$, $|c|^2 + |d|^2 =1$, and
\begin{equation}
|\psi(0)\rangle = \bigg{[} a |g,2\rangle^{(1)} +  b |e,0\rangle^{(1)}\bigg{]} \bigg{[} c |g,2\rangle^{(2)} +  d |e,0\rangle^{(2)}\bigg{]}\times
\end{equation}
\[
\times\bigg{[} e|g,2\rangle^{(3)} +  f |e,0\rangle^{(3)}\bigg{]},
\]
where $|a|^2 + |b|^2 =1$, $|c|^2 + |d|^2 =1$, and $|e|^2 + |f|^2 =1$.

The initial state (19) is symmetric under the cavity interchange $2\leftrightarrow 3$ and so $|\psi(t)\rangle$ always remains symmetrical under the interchange $2\leftrightarrow 3$ owing to the Hamiltonian (1) being symmetrical. The initial state (20) is strictly asymmetric. The initial state (21) may be symmetric or asymmetric under the interchange of any two or all three cavities depending on the values of the initial amplitudes. For instance, if $a=c$ and $b=d$, then the solution of the Schr\"{o}dinger equation  will be symmetric in the interchange $1\leftrightarrow 2$ for all times. If, however, $a=c=e$ and $b=d=f$, then the solution is symmetric in the interchanges  of all three cavities. This latter case is considered in Appendix D.

In the limit of large hopping strength, $\hbar \xi \gg [(E^{+}_{4} - E^{-}_{4}) \cos^2 \theta_{4}]$, the vectors in the Hilbert space associated with the sectors $a$, $b$, $c$, and $d$ are uncoupled since the hopping Hamiltonian in (1) only exchanges photons and not atomic excitations. The equations for the 38 probability amplitudes break up into linear differential equations involving ten, eighteen, nine, and one amplitudes, respectively [see (C2), (C3), (C8), (C9), (C11), and (C13)]. Note that for the case when all three cavities are in the excited state, the probability amplitude remains constant in time [see (C13)]. All these differential equations can be solved analytically in terms of exponential functions for arbitrary initial states.

\subsection{ Initial state Eq. (19)}

The initial state given by Eq. (19) lies in the non-overlapping sectors $a$ and $b$ in the large hopping limit $\hbar \xi \gg [(E^{+}_{4} - E^{-}_{4}) \cos^2 \theta_{4}]$ presented in Appendix C. The time development of the system is given by a linear superposition of vectors in these two disjoint sectors. We consider the initial state with $a=1$ and so $b=0$, that is, $A(0)=1$ in (C1) with the solution in the subspace of sector $a$. The solution is given by (C4), where all the probability amplitudes have analytic solutions in terms of exponential functions and the results for $A(t)$ and $F(t)$ are given explicitly by (C5) and (C6), respectively. Maximal entanglement is obtained by minimizing $|A(t)|^2 + |F(t)|^2$, which occurs for $\xi \tau \approx 1.7500$, where $|A(\tau)|^2 + |F(\tau)|^2 \approx 0.001833$, $2|B(\tau)|^2 \approx 0.140493$, $2|E(\tau)|^2\approx 0.055394$, $2|G(\tau)|^2 \approx 0.459478$, and $2|K(\tau)|^2 \approx 0.342801$. This gives a measure of entanglement $ \log_{2} (546) \approx 9.1$.

On the other hand, one finds a maximum in $|A(t)|^2 + |F(t)|^2$, which occurs for $\xi \tau \approx 3.0318$, where $|A(\tau)|^2 + |F(\tau)|^2 \approx 0.95166$, $|A(\tau)|^2 \approx 0.89530$, $2|B(\tau)|^2 \approx 0.00137$, $2|E(\tau)|^2 \approx 0.02296$, $|F(\tau)|^2 \approx 0.05637$,  $2|G(\tau)|^2 \approx 0.02225$, and $2|K(\tau)|^2 \approx 0.00176$. The system does not return to its initial state owing to the eigenfrequencies being incommensurate.

\subsection{ Initial state Eq. (20)}

The initial state given by Eq. (20) lies in the non-overlapping sectors $a$, $b$, and $c$ in the large hopping limit $\hbar \xi \gg [(E^{+}_{4} - E^{-}_{4}) \cos^2 \theta_{4}]$ presented in Appendix C. We consider the case where the amplitudes $a=0$, $b=c=1$, and $d=0$ in Eq. (20), that is, the initial unentangled strictly asymmetric state $ |\psi(0)\rangle =|e,2\rangle^{(1)}|g,2\rangle^{(2)}|g,0\rangle^{(3)}$, that is, $D(0)=1$ in (C7), which lies in sector $b$. The state of the system is given for later times by the entangled state
\[
|\psi(t)\rangle = A(t) \Big{[}|e,0\rangle^{(1)}|g,2\rangle^{(2)}  + |e,2\rangle^{(1)} |g,0\rangle^{(2)} \Big{]}|g,2\rangle^{(3)}
\]
\begin{equation}
+ B(t) \Big{[} |e,0\rangle^{(1)} |g,4\rangle^{(2)} +  |e,4\rangle^{(1)} |g,0\rangle^{(2)}\Big{]}|g,0\rangle^{(3)}
\end{equation}
\[
+ C(t) |e,0\rangle^{(1)}|g,0\rangle^{(2)}|g,4\rangle^{(3)} + D(t) |e,2\rangle^{(1)} |g,2\rangle^{(2)}|g,0\rangle^{(3)},
\]
where the probability amplitudes in (22) are given in Eq. (E3). Note that the system returns to its initial state for times $\tau$ such that $\xi \tau= n \pi$ for $n=1, 2, 3, \cdots$ and that $D(t) \neq 0$ for $t$ real. In addition, for any probability amplitude $X(\xi t)$, $|X(\xi t)|^2 = |X(\pi - \xi t)|^2 $.

Maximal entanglement is obtained by minimizing $|C(t)|^2 + |D(t)|^2$, which cannot vanish owing to the initial state being strictly asymmetric. Minimization occurs for $\xi \tau \approx 0.1930, 0.8542$, where the minima have the same values and is $|C(\tau)|^2 + |D(\tau)|^2 \approx 0.1829$. The measure of  entanglement is $\log_{2}(5.467) \approx 2.450$.

There are several simpler tripartite entangled states that occur at different times when some probability amplitudes vanish: (a) $\xi T_{1}= \pi/5$, $B(T_{1})=0$, $C(T_{1})=0$, $|A(T_{1})|^2 = \frac{4}{9} \sin^2(\pi/5) \approx 0.1536$ and $|D(T_{1})|^2 = 1-\frac{8}{9}\sin^2(\pi/5) \approx 0.6929$ and so $2|A(T_{1})|^2 + |D(T_{1})|^2 =1$; (b) $\xi T_{2}= \pi/3 $, $A(T_{2})=0$, $B(T_{2})=0$, $|C(T_{2})|^2 = 18/25$ and $|D(T_{2})|^2 = 7/25$ with $|C(T_{2})|^2 + |D(T_{2})|^2 =1$; and (c) $\xi T_{3}= \frac{\pi}{2} -\frac{1}{2}\arccos([\sqrt{5}+1]/4) \approx 1.2566$, $B(T_{3})=0$, $C(T_{3})=0$, $|A(T_{3})|^2 = \frac{2}{9}[ 1+ \cos(\pi/5)] \approx 0.4020$ and $|D(T_{2})|^2 = \frac{5}{9} -\frac{4}{9} \cos(\pi/5) \approx 0.1960$ with $2|A(T_{1})|^2 + |D(T_{1})|^2 =1$.

Is is interesting that for the above case with $\xi \tau =\pi/3$, one has that $A(\tau)=B(\tau) =0$ with $C(\pi/3) =\sqrt{6}(3-\sqrt{3}i)/10$ and $D(\pi/3)= (2+ \sqrt{3} i)/5$. Therefore, the probability that the system is in its initial state with two photons each in cavities 1 and 2 is $28\%$ and that all four photons are in cavity 3 is $72\%$.

\subsection{ Initial state Eq. (21)}

In Appendix D, the explicit solution is given for an initially unentangled, symmetric state in all three cavities, which can occur only in the $N=6$ manifold, viz., the initial state given by  Eq. (21) with $a=c=e$ and $b=d=f$. Solutions (D8)--(D11) are not periodic owing to the incommensurate nature of the eigenfrequencies. However, for the initial state
$|g,2\rangle^{(1)}|g,2\rangle^{(2)}|g,2\rangle^{(3)}$, the solution is given by (D8), which are periodic with period $\tau= \pi/\sqrt{66}\xi$ and so
\[
|\psi(t)\rangle = A(t) |g,2\rangle^{(1)}|g,2\rangle^{(2)}|g,2\rangle^{(3)}
\]
\begin{equation}
+  F(t) \frac{1}{\sqrt{6}} \sum_{P} P|g,4\rangle^{(1)}|g,2\rangle^{(2)}|g,0\rangle^{(3)}
\end{equation}
\[
+ K(t)  \frac{1}{\sqrt{3}} \sum_{P} P |g,6\rangle^{(1)}|g,0\rangle^{(2)}|g,0\rangle^{(3)},
\]
where the probability amplitudes $A(t)$, $F(t)$, and $K(t)$ are given by (D8) with $a=1$.

The amplitude $A(t)$ vanishes for $\cos(2 \sqrt{66} \xi T ) =-5/6$ resulting in the maximally entangled state
\[
|\psi(T)\rangle = \pm \frac{i}{\sqrt{6}} \frac{1}{\sqrt{6}} \sum_{P} P|g,4\rangle^{(1)}|g,2\rangle^{(2)}|g,0\rangle^{(3)}
\]
\begin{equation}
- \frac{\sqrt{5}}{\sqrt{6}}  \frac{1}{\sqrt{3}} \sum_{P} P |g,6\rangle^{(1)}|g,0\rangle^{(2)}|g,0\rangle^{(3)}.
\end{equation}

Note that if $K(\tau)=0$, that is, $\xi \tau= \pi l/\sqrt{66}$, $l=0, 1, 2, \cdots$, then $F(\tau) =0$ according to (D8); however, the converse does not follow. Therefore, for times $\xi \tau=(l+1/2)\pi/\sqrt{66}$, $l=0, 1, 2, \cdots$, $F(\tau)=0$ but $K(\tau) = -2\sqrt{30}/11$, and so
\[
|\psi(\tau)\rangle =- \frac{1}{11} |g,2\rangle^{(1)}|g,2\rangle^{(2)}|g,2\rangle^{(3)}
\]
\begin{equation}
-\frac{2 \sqrt{30}}{11}   \frac{1}{\sqrt{3}} \sum_{P} P |g,6\rangle^{(1)}|g,0\rangle^{(2)}|g,0\rangle^{(3)}.
\end{equation}
The geometric measure of entanglement of the state (25) is $E(|\psi(\tau)\rangle) = \log_{2} 121 \approx 6.92$. This type of entanglement encoding is that of a W-state, a tripartite state of three qubits \cite{DVC00}. On the other hand, if $\tau=(l+1/2)\pi/2\sqrt{66}\xi$, $l=0, 1, 2, \cdots$, then
\[
|\psi(\tau)\rangle = \frac{5}{11} |g,2\rangle^{(1)}|g,2\rangle^{(2)}|g,2\rangle^{(3)}
\]
\begin{equation}
+ \frac{\sqrt{66} i}{11} (-1)^l \frac{1}{\sqrt{6}} \sum_{P}  P|g,4\rangle^{(1)}|g,2\rangle^{(2)}|g,0\rangle^{(3)}
\end{equation}
\[
 - \frac{\sqrt{30}}{11}  \frac{1}{\sqrt{3}} \sum_{P} P |g,6\rangle^{(1)}|g,0\rangle^{(2)}|g,0\rangle^{(3)}.
\]
The geometric measure of entanglement of the state (26) is $E(|\psi(\tau)\rangle) = \log_{2} \frac{121}{25} \approx 2.28$.

\section{summary and conclusions}

We have studied the dynamical behavior of a deterministic system constituted by three identical cavities each enclosing a three-level atom with intracavity interactions governed by two-coherent-photon hopping.  We consider atom/photon states for each cavity that corresponds to a multileveled system of three-, five-, and seven-dimensional spaces. However, the dynamical state of the system, owing to conservation laws, lie in subspaces given by 6, 18, and 38 dimensions rather than the usual $3^3 =27$, $5^3=125$, and $7^3=343$, respectively.  Explicit analytic solutions are found for arbitrary initial unentangled pure states in the manifolds $N$=2, 4, and 6 when the exchange of photons between cavities occurs at a much faster rate than the rate of atomic transitions, the large hopping limit. Tripartite entanglement between the cavities can be used to generate bipartite entanglement between any two cavities via local quantum operations, which can be performed either on the atom or on the photon pairs or even on both the atom/photon state of a given cavity. Dynamically generated tripartite entangled states, which reflect the symmetry of the Hamiltonian and the symmetry properties of the initially unentangled state, are given by the superposition of symmetric, antisymmetric, and asymmetric states. The symmetric states are generalizations of the W-states.

\appendix

\section{N=2 manifold}

The Hilbert space for three cavities with the state of each cavity given by the qutrit $|g,0\rangle$, $|g,2\rangle$, and $|e,0\rangle$, normally lies in a Hilbert space of $3^3 =27$ dimensions. However, the dynamics of the $N=2$ manifold is governed by only a 6-dimensional subspace owing to the constancy of the operator $\hat{N}$ with general state vector
\[
|\psi(t)\rangle = A(t) |g,0\rangle^{(1)} |g,0\rangle^{(2)}|g,2\rangle^{(3)}
\]
\begin{equation}
+B(t) |g,0\rangle^{(1)} |g,2\rangle^{(2)}|g,0\rangle^{(3)}+C(t) |g,2\rangle^{(1)} |g,0\rangle^{(2)}|g,0\rangle^{(3)}
\end{equation}
\[
+ D(t) |g,0\rangle^{(1)} |g,0\rangle^{(2)}|e,0\rangle^{(3)}+E(t) |g,0\rangle^{(1)} |e,0\rangle^{(2)}|g,0\rangle^{(3)}
\]
\[
+ F(t) |e,0\rangle^{(1)} |g,0\rangle^{(2)}|g,0\rangle^{(3)}.
\]
The dynamical equations of motion are
\[
i\dot{A} = A + 2 \xi B + 2 \xi C + \tan\theta_{0} D,
\]
\begin{equation}
i\dot{B} = 2 \xi A +  B + 2 \xi C + \tan\theta_{0} E,
\end{equation}
\[
i\dot{C} = 2 \xi A + 2 \xi B +C + \tan\theta_{0} F,
\]
\[
\dot{D} =  \tan\theta_{0} A  + \tan^2\theta_{0} D,
\]
\[
\dot{E} =  \tan\theta_{0} B  + \tan^2\theta_{0} E,
\]
\[
\dot{F} =  \tan\theta_{0} C  + \tan^2\theta_{0} F,
\]
where we have introduced the dimensionless time $[(E_{0}^{+} - E_{0}^{-}) \cos^2 \theta_{0}] t/\hbar \rightarrow t $ and the dimensionless hopping coupling $ \hbar\xi/[ (E_{0}^{+} - E_{0}^{-}) \cos^2 \theta_{0} ] \rightarrow \xi$.

In the limit of large hopping strength $\xi\gg 1$, the solutions are
\[
A(t) \approx \frac{1}{3} \Big{[} A(0)+B(0)+C(0) \Big{]} e^{-4i \xi t}
\]
\[
+ \frac{1}{3} \Big{[} 2 A(0) -B(0) -C(0)\Big{]} e^{2i \xi t},
\]
\[
B(t)\approx \frac{1}{3} \Big{[} A(0)+B(0)+C(0) \Big{]} e^{-4i \xi t}
\]
\begin{equation}
+  \frac{1}{3} \Big{[} - A(0)+ 2 B(0) -C(0) \Big{]} e^{2i \xi t},
\end{equation}
\[
C(t)\approx \frac{1}{3} \Big{[} A(0)+B(0)+C(0) \Big{]} e^{-4i \xi t}
\]
\[
+  \frac{1}{3} \Big{[} - A(0) - B(0)  +2C(0) \Big{]} e^{2i \xi t},
\]
\[
D(t) \approx D(0) \hspace{0.2in}  E(t) \approx E(0) \hspace{0.2in} F(t) \approx F(0),
\]
where the amplitudes are periodic with period $T= \pi / \xi $. Results (9) in the text follow from (A1) and (A3) for $A(0) =a$, $B(0)=0$, $C(0)=0$, $D(0)=b$, $E(0)=0$, and $F(0)=0$. Note the difference in the labeling of the probability amplitudes in (A1) and Eq. (6). The average time the photons spend in each cavity depends on the initial state of the system and is bounded for cavity 3 by $ \frac{1}{\pi} \int_{0}^{\pi} |A(t)|^2 \leq \frac{2}{9} + \frac{1}{3} |A(0)|^2$ with analogous bounds for the amplitudes $B(t)$ and $C(t)$, for cavities 2 and 1, respectively. The equality holds when the photons are initially in cavity 3, that is, $A(0)=1$, in which case the photons spend $5/9$ of the time in cavity 3 and $2/9$ each in cavities 1 and 2.

\section{N=4 manifold}

The Hilbert space for three cavities with the state of each cavity given by the five-dimensional qudit $|g,4\rangle$, $|g,2\rangle$, $|g,0\rangle$, $|e,0\rangle$, and $|e,2\rangle$, normally lies in a Hilbert space of $5^3 =125$ dimensions. However, the dynamics of the $N=4$ manifold is governed by only an 18-dimensional subspace with general state vector
\[
|\psi(t)\rangle = A(t) |g,4\rangle^{(1)} |g,0\rangle^{(2)}|g,0\rangle^{(3)}
\]
\[
+ |g,2\rangle^{(1)} \Bigg{[} B(t) |g,0\rangle^{(2)}|g,2\rangle^{(3)}  +  C(t)  |g,2\rangle^{(2)} |g,0\rangle^{(3)}
\]
\[
+D(t)|e,0\rangle^{(2)}|g,0\rangle^{(3)}+ E(t) |g,0\rangle^{(2)}|e,0\rangle^{(3)} \Bigg{]}
\]
\[
+ |g,0\rangle^{(1)} \Bigg{[} F(t) |g,0\rangle^{(2)}|g,4\rangle^{(3)} +G(t) |g,4\rangle^{(2)}|g,0\rangle^{(3)}
 \]
 \[
+ J(t) |e,2\rangle^{(2)}|g,0\rangle^{(3)} + K(t) |g,0\rangle^{(2)}|e,2\rangle^{(3)}
\]
\begin{equation}
+ L(t) |e,0\rangle^{(2)}|e,0\rangle^{(3)}+M(t) |e,0\rangle^{(2)}|g,2\rangle^{(3)}
\end{equation}
\[
+ N(t) |g,2\rangle^{(2)}|e,0\rangle^{(3)} + P(t) |g,2\rangle^{(2)}|g,2\rangle^{(3)}\Bigg{]}
\]
\[
+ |e,0\rangle^{(1)} \Bigg{[}  R(t) |g,0\rangle^{(2)}|g,2\rangle^{(3)} + S(t) |g,2\rangle^{(2)}|g,0\rangle^{(3)}
\]
\[
+T(t) |e,0\rangle^{(2)}|g,0\rangle^{(3)} + U(t) |g,0\rangle^{(2)}|e,0\rangle^{(3)} \Bigg{]}
\]
\[
 + W(t) |e,2\rangle^{(1)}  |g,0\rangle^{(2)}|g,0\rangle^{(3)}.
\]
The dynamical equations for the probabilities amplitudes in (B1), in the limit of large hopping strength $\hbar \xi \gg [(E^{+}_{2} - E^{-}_{2}) \cos^2 \theta_{2}]$, are given by the following uncoupled sets: For the amplitudes $A$, $B$, $C$, $F$, $G$ and, $P$
\[
i \dot{A} \approx \sqrt{24} \xi ( B + C),
\]
\[
i (\dot{B} +\dot{C}) \approx  2\sqrt{24} \xi A + 2 \xi(B+ C )
\]
\begin{equation}
+ \sqrt{24} \xi (G+F ) + 4 \xi P,
\end{equation}
\[
i (\dot{G}+ \dot{F} )\approx \sqrt{24} \xi (B+C)  + 2 \sqrt{24} \xi P,
\]
\[
i \dot{P}  \approx   2 \xi (B +C) + \sqrt{24} \xi (G+F),
\]
and
\begin{equation}
i (\dot{B} -\dot{C}) \approx   - 2 \xi(B - C ) - \sqrt{24} \xi (G-F ).
\end{equation}
\[
i (\dot{G}- \dot{F} )\approx - \sqrt{24} \xi (B-C).
\]
the eigenfrequencies for the system of equations (B2) and (B3) are $-8\xi$, $-6\xi$, $4\xi$, and $12\xi$.

For the amplitudes $D$, $J$, and $M$ and
\[
i \dot{D}\approx 2 \xi J + 2 \xi M,
\]
\begin{equation}
i \dot{J} \approx 2 \xi D + 2 \xi M,
\end{equation}
\[
i \dot{M}  \approx 2 \xi D + 2 \xi J.
\]

For the amplitudes $E$, $K$, and $N$
\[
i \dot{E}  \approx 2 \xi K + 2 \xi N,
\]
\begin{equation}
i \dot{K}  \approx 2 \xi E + 2 \xi N,
\end{equation}
\[
i  \dot{N}  \approx 2 \xi E + 2 \xi K.
\]

For the amplitudes $L$, $T$, and $U$
\begin{equation}
\dot{L} \approx \dot {T} \approx  \dot{U} \approx 0.
\end{equation}

Finally, for the amplitudes $R$, $S$, and $W$
\[
i \dot{R}  \approx 2 \xi S + 2 \xi W,
\]
\begin{equation}
i  \dot{S} \approx 2 \xi R + 2 \xi W,
\end{equation}
\[
i \dot{W}\approx 2 \xi R + 2 \xi S.
\]
Solutions of the system of equations given by (B4), (B5), and (B7) are the same as solution (A3) for the system of equations (A2) in the limit $\xi\gg 1$. The eigenfrequencies for the Eqs. (B4), (B5), and (B7) are -$2 \xi$, and $4 \xi$.

\section{N=6 manifold}

The Hilbert space for three cavities with the state of each cavity given by the seven-dimensional qudit $|g,6\rangle$, $|g,4\rangle$, $|g,2\rangle$, $|g,0\rangle$, $|e,0\rangle$,  $|e,2\rangle$, and $|e,4\rangle$, normally lies in a Hilbert space of $7^3 =343$ dimensions. However, in the limit of large hopping strength $\hbar \xi \gg [(E^{+}_{4} - E^{-}_{4}) \cos^2 \theta_{4}]$ the dynamics of the $N=6$ manifold is governed by only a 38-dimensional subspace that in the limit of large hopping strength separates into four sectors: (a) all three cavities are in the ground state, (b) one cavity is in the excited state, (c) two cavities are in the excited state, and (d) all three cavities are in the excited state.

\subsection {Sector a}

The general state when all three cavities are in the ground state is
\[
|\psi(t)\rangle = A(t) |g,6\rangle^{(1)}|g,0\rangle^{(2)}|g,0\rangle^{(3)}
\]
\[
+ B(t) |g,4\rangle^{(1)}|g,2\rangle^{(2)}|g,0\rangle^{(3)}+ C(t) |g,4\rangle^{(1)}|g,0\rangle^{(2)}|g,2\rangle^{(3)}
\]
\begin{equation}
+ D(t)|g,2\rangle^{(1)}|g,4\rangle^{(2)}|g,0\rangle^{(3)}+E(t) |g,2\rangle^{(1)}|g,0\rangle^{(2)}|g,4\rangle^{(3)}
\end{equation}
\[
+ F(t) |g,2\rangle^{(1)}|g,2\rangle^{(2)}|g,2\rangle^{(3)}+G(t) |g,0\rangle^{(1)}|g,6\rangle^{(2)}|g,0\rangle^{(3)}
\]
\[
+ J(t) |g,0\rangle^{(1)}|g,4\rangle^{(2)}|g,2\rangle^{(3)}+K(t) |g,0\rangle^{(1)}|g,2\rangle^{(2)}|g,4\rangle^{(3)}
\]
\[
+L(t) |g,0\rangle^{(1)}|g,0\rangle^{(2)}|g,6\rangle^{(3)}.
\]
The equations governing the time behavior of the probability amplitudes break up into two groups,
\[
i (\dot{B} -\dot{C}) \approx 2 \xi (C-B) +12 \xi (D-E)
\]
\[
i (\dot{D} -\dot{E})  \approx 12 \xi (B-C) + \sqrt{60} \xi (G-L) +2 \xi (J-K)
\]
\begin{equation}
i (\dot{G} -\dot{L}) \approx \sqrt{60} \xi (D-E+J-K)
\end{equation}
\[
i (\dot{J} -\dot{K}) \approx 2 \xi (D-E)+ \sqrt{60} \xi (G-L) +12 \xi (K-J),
\]
with eigenfrequencies $14 \xi$, $-2\xi$, and $(1 \pm \sqrt{241})\xi$; and
\[
i (\dot{B} +\dot{C}) \approx 2 \sqrt{60}  \xi A + 2 \xi (B+C) +12 \xi (D+E) +2\sqrt{24} \xi F
\]
\[
i (\dot{D} +\dot{E}) \approx 12  \xi (B+C) + \sqrt{60} \xi (G+L) +2 \xi (J+K) +2\sqrt{24} \xi F
\]
\begin{equation}
i (\dot{G} +\dot{L})  \approx \sqrt{60} \xi (D+E+J+K)
\end{equation}
\[
i (\dot{J} +\dot{K}) \approx 2 \xi (D+E)+ \sqrt{60} \xi (G+L) +12  \xi (J+K) +2\sqrt{24} \xi F
\]
\[
i \dot{F} \approx  \sqrt{24} \xi (B+C +D+E+J+K)
\]
\[
i\dot{A} \approx \sqrt{60} \xi (B+C),
\]
with eigenfrequencies 0, $2\xi$, $(-1\pm\sqrt{241})\xi$, and $(7 \pm\sqrt{313})\xi$. Note that the eigenfrequencies are incommensurable and so the system never returns to its initial state albeit it can get arbitrarily close to it.

Consider the case with initial condition $A(0)=1$ in (C1). Eq. (C2) implies that $B(t)=C(t)$, $D(t)=E(t)$,  $G(t)= L(t)$, and $J(t)=K(t)$ and so

\[
|\psi(t)\rangle = A(t) |g,6\rangle^{(1)}|g,0\rangle^{(2)}|g,0\rangle^{(3)}
\]
\[
+ B(t) |g,4\rangle^{(1)} \bigg{[}|g,2\rangle^{(2)}|g,0\rangle^{(3)}+  |g,0\rangle^{(2)}|g,2\rangle^{(3)}\bigg{]}
\]
\begin{equation}
+ E(t)|g,2\rangle^{(1)} \bigg{[}|g,4\rangle^{(2)}|g,0\rangle^{(3)}+|g,0\rangle^{(2)}|g,4\rangle^{(3)}\bigg{]}
\end{equation}
\[
+ G(t) |g,0\rangle^{(1)} \bigg{[}|g,6\rangle^{(2)}|g,0\rangle^{(3)}+ |g,0\rangle^{(2)}|g,6\rangle^{(3)}\bigg{]}
\]
\[
+ K(t) |g,0\rangle^{(1)} \bigg{[}|g,4\rangle^{(2)}|g,2\rangle^{(3)}+ |g,2\rangle^{(2)}|g,4\rangle^{(3)}\bigg{]}
\]
\[
+F(t) |g,2\rangle^{(1)}|g,2\rangle^{(2)}|g,2\rangle^{(3)},
\]
with normalization $|A(t)|^2  +2 \big{(}|B(t)|^2 + |E(t)|^2 + G(t)|^2 +|K(t)|^2 \big{)} + |F(t)|^2=1$

The system of equations (C3) can be solved explicitly, in particular, the probability amplitudes $A(t)$  and $F(t)$ for the unentangled states in  (C4) are given by
\[
A(t)= \frac{2}{11} + \frac{10}{29} e^{-2i\xi t} +\frac{5}{66} \big{(} 1 + \frac{7}{\sqrt{313}} \big{)} e^{(-7  + \sqrt{313})i \xi t}
\]
\begin{equation}
+ \frac{5}{66} \big{(} 1  - \frac{7}{\sqrt{313}} \big{)} e^{-(7  + \sqrt{313})i \xi t} + \frac{14}{87} \big{(} 1 + \frac{8} {7 \sqrt{241}} \big{)}  e^{ (1 +\sqrt{241}) i \xi t}
\end{equation}
\[
+\frac{14}{87} \big{(} 1 - \frac{8} {7 \sqrt{241}} \big{)}  e^{ (1 -\sqrt{241}) i \xi t}
\]
and
\[
F(t) = -\frac{\sqrt{10}}{11} + \frac{\sqrt{10}}{22} \bigg{(} 1 + \frac{7}{\sqrt{313}} \bigg{)} e^{-(7 - \sqrt{313})i \xi t}
\]
\begin{equation}
+ \frac{\sqrt{10}}{22} \bigg{(} 1- \frac{7}{\sqrt{313}}\bigg{)} e^{-(7 + \sqrt{313})i \xi t}.
\end{equation}

\subsection {Sector b}

The general state for one of the cavities to be in the excited state is
\[
|\psi(t)\rangle = A(t) |e,0\rangle^{(1)}|g,2\rangle^{(2)}|g,2\rangle^{(3)}
\]
\begin{equation}
+ B(t) |e,0\rangle^{(1)}|g,4\rangle^{(2)}|g,0\rangle^{(3)}+ C(t) |e,0\rangle^{(1)}|g,0\rangle^{(2)}|g,4\rangle^{(3)}
\end{equation}
\[
+ D(t) |e,2\rangle^{(1)}|g,2\rangle^{(2)}|g,0\rangle^{(3)}+E(t) |e,2\rangle^{(1)}|g,0\rangle^{(2)}|g,2\rangle^{(3)}
\]
\[
+  F(t) |e,4\rangle^{(1)}|g,0\rangle^{(2)}|g,0\rangle^{(3)}.
\]
The equations governing the time behavior of the probability amplitudes break up into two groups,
\[
i (\dot{A} -\dot{D}) \approx 2 \xi (D-A) + \sqrt{24} \xi (C-F)
\]
\begin{equation}
i (\dot{C} -\dot{F}) \approx  \sqrt{24} \xi (A-D),
\end{equation}
with eigenfrequencies $-6\xi$ and $4\xi$,
and
\[
i (\dot{A} + \dot{D}) \approx 2 \xi ( A+D)  + 4 \xi E + \sqrt{24} \xi ( C +F) + 2\sqrt{24} \xi B
\]
\begin{equation}
i (\dot{C} + \dot{F}) \approx \sqrt{24} \xi (A+D)  + 2 \sqrt{24} \xi E
\end{equation}
\[
i \dot{B} \approx  \sqrt{24} \xi (A  + D)
\]
\[
i \dot{E} \approx  2 \xi (A  + D) + \sqrt{24} \xi (C+F),
\]
with eigenfrequencies $-8\xi$, $-6\xi$, $4\xi$, and $12\xi$.

\subsection {Sector c}

The general state for two cavities to be in the excited state is
\[
|\psi(t)\rangle = A(t) |e,2\rangle^{(1)}|e,0\rangle^{(2)}|g,0\rangle^{(3)}
\]
\begin{equation}
+ B(t) |e,0\rangle^{(1)}|e,2\rangle^{(2)}|g,0\rangle^{(3)}+ C(t) |e,0\rangle^{(1)}|e,0\rangle^{(2)}|g,2\rangle^{(3)}.
\end{equation}
The equations governing the time behavior of the probability amplitudes are
\[
i \dot{A}  \approx 2 \xi B + 2 \xi C,
\]
\begin{equation}
i  \dot{B} \approx 2 \xi A + 2 \xi C,
\end{equation}
\[
i \dot{C}\approx 2 \xi A + 2 \xi B.
\]
Solutions of the system of equations given by (C8) are the same as solution (A3) for the system of equations (A2).

\subsection {Sector d}
The general state for three cavities to be in the excited state is
\begin{equation}
|\psi(t)\rangle = A(t) |e,0\rangle^{(1)}|e,0\rangle^{(2)}|e,0\rangle^{(3)}.
\end{equation}
The equation governing the time development is given by
\begin{equation}
\dot{A}\approx 0
\end{equation}
and so $A(t)\approx A(0)$.

\section{Symmetric N=6}

One can obtain the 38-dimensional vector space for the $N=6$ manifold from (A1) by applying the following replacements: $|g,4\rangle^{(1)}$ by $|g,6\rangle^{(1)}$, $|g,2\rangle^{(1)}$ by a linear combination of the vectors $|g,4\rangle^{(1)}$ and  $|e,2\rangle^{(1)}$,  $|g,0\rangle^{(1)}$ by a linear combination of the vectors $|g,2\rangle^{(1)}$ and  $|e,0\rangle^{(1)}$, $|e,0\rangle^{(1)}$ by a linear combination of the vectors $|g,4\rangle^{(1)}$ and  $|e,2\rangle^{(1)}$, and  $|e,2\rangle^{(1)}$ by  $|e,4\rangle^{(1)}$. The preceding gives rise to 26 vectors. In addition, one must add to that set the 10 vectors that are the product of the vector  $|g,0\rangle^{(1)}$  and the linear superposition of the vectors $|g,6\rangle^{(2)}|g,0\rangle^{(3)}$, $|g,0\rangle^{(2)}|g,6\rangle^{(3)}$, $|g,4\rangle^{(2)}|e,0\rangle^{(3)}$, $|e,0\rangle^{(2)}|g,4\rangle^{(3)}$, $|g,4\rangle^{(2)}|g,2\rangle^{(3)}$, $|g,2\rangle^{(2)}|g,4\rangle^{(3)}$, $|e,2\rangle^{(2)}|g,2\rangle^{(3)}$, $|g,2\rangle^{(2)}|e,2\rangle^{(3)}$, $|e,4\rangle^{(2)}|g,0\rangle^{(3)}$, and $|g,0\rangle^{(2)}|e,4\rangle^{(3)}$. Finally, we have to add unentangled symmetric vectors, which cannot occur in the $N=2$ and $N=4$ manifolds, viz.,  $|g,2\rangle^{(1)}|g,2\rangle^{(2)}|g,2\rangle^{(3)}$ and $|e,0\rangle^{(1)}|e,0\rangle^{(2)}|e,0\rangle^{(3)}$, for a total of 38 vectors.

In the $N=6$ manifold, the initial unentangled state that is totally symmetric in all three cavities is given by
\[
|\psi(0)\rangle = \Big{[} a|g,2\rangle^{(1)}  + b |e,0\rangle^{(1)}\Big{]}  \Big{[} a|g,2\rangle^{(2)}  + b|e,0\rangle^{(2)} \Big{]} \times
\]
\begin{equation}
\times\Big{[}a |g,2\rangle^{(3)}  + b |e,0\rangle^{(3)}\Big{]},
\end{equation}
where $|a|^2  + |b|^2=1$.  The case with $b=0$ reduces to that considered in Appendix C in sector $a$ for $\xi\gg 1$. The trivial case $a=0$ gives a constant amplitude as indicated in (C13).

The general symmetric state is given by
\[
|\psi(t)\rangle = A(t) |g,2\rangle^{(1)}|g,2\rangle^{(2)}|g,2\rangle^{(3)}
\]
\[
+ B(t) \frac{1}{\sqrt{3}} \sum_{P}  P |e,0\rangle^{(1)}|g,2\rangle^{(2)}|g,2\rangle^{(3)}
\]
\[
+  C(t) \frac{1}{\sqrt{3}} \sum_{P} P|e,0\rangle^{(1)}|e,0\rangle^{(2)}|g,2\rangle^{(3)}
\]
\[
+ D(t) |e,0\rangle^{(1)}|e,0\rangle^{(2)}|e,0\rangle^{(3)}
\]
\begin{equation}
+ E(t) \frac{1}{\sqrt{6}} \sum_{P} P |g,4\rangle^{(1)}|e,0\rangle^{(2)}|g,0\rangle^{(3)}
\end{equation}
\[
+  F(t) \frac{1}{\sqrt{6}} \sum_{P} P|g,4\rangle^{(1)}|g,2\rangle^{(2)}|g,0\rangle^{(3)}
\]
\[
+ G(t) \frac{1}{\sqrt{6}} \sum_{P} P |e,2\rangle^{(1)}|g,2\rangle^{(2)}|g,0\rangle^{(3)}
\]
\[
+ H(t) \frac{1}{\sqrt{6}}  \sum_{P}P  |e,2\rangle^{(1)}|e,0\rangle^{(2)}|g,0\rangle^{(3)}
\]
\[
+ K(t)  \frac{1}{\sqrt{3}} \sum_{P} P |g,6\rangle^{(1)}|g,0\rangle^{(2)}|g,0\rangle^{(3)}
\]
\[
+ J(t) \frac{1}{\sqrt{3}} \sum_{P} P |e,4\rangle^{(1)}|g,0\rangle^{(2)}|g,0\rangle^{(3)},
\]
where for the amplitudes $B(t)$, $C(t)$, $K(t)$, and $J(t)$ the sum is only over even permutation of the cavities; whereas, for $E(t)$, $F(t)$, $G(t)$, and $H(t)$ the sum is over both even and odd permutation. The state (D2) brings forth all 38 states that span the $N=6$ manifold, which corresponds to 10 symmetric, 4 antisymmetric, and 24 asymmetric states that treats all the three cavities on the same footing. The antisymmetric states follow from the states associated with the amplitudes $E(t)$, $F(t)$, $G(t)$, and $H(t)$ in (D2), where a $-1$ is inserted before those terms arising from an odd permutation.

The symmetric states are either entangled or unentangled, with the former coming in two varieties, viz., involving three or six states.  The presence of only two unentangled symmetric states indicates that in the $N=6$ manifold one does not have the general GHZ maximally entangled state for the seven-dimensional qudit associated with each of the three cavities in the space spanned by $7^3= 343$ vectors,
 \begin{equation}
 |GHZ\rangle = \frac{1}{\sqrt{7}}\sum_{j} |j \rangle^{(1)}|j \rangle^{(2)}|j \rangle^{(3)},
 \end{equation}
where the sum over $j$ is over the states $|g,6\rangle$, $|g,4\rangle$, $|g,2\rangle$, $|g,0\rangle$, $|e,0\rangle$, $|e,2\rangle$, and $|e,4\rangle$ and the trace over one of the three cavities gives rise to an unentangled mixed state. In the $N=6$ manifold, the trace over one of the three cavities of the symmetric states involving six states produces an entangled reduced density matrix for each cavity given by the three states that span that space. However, a trace over two of the three cavities, for instance, the reduced density matrix for each cavity associated with the probability amplitude $E(t)$ in (D2) is given by $\rho = \frac{1}{3} [ |g,0\rangle \langle g,0| +|e,0\rangle \langle e,0|+|g,4\rangle \langle g,4|$.

In the large hopping limit $\xi \gg [(E^{+}_{4} - E^{-}_{4}) \cos^2 \theta_{4}]$ the equations for the probability amplitudes decouple and are as follows: For $A(t)$, $F(t)$ and $K(t)$
\[
i \frac{d A}{dt} \approx 12 \xi F,
\]
\begin{equation}
i \frac{d F}{dt} \approx 12 \xi A  + 2\sqrt{30} \xi K,
\end{equation}
\[
i \frac{d K}{dt} \approx 2 \sqrt{30} \xi F,
\]
for  $B(t)$, $E(t)$, $G(t)$, and $J(t)$
\[
i \frac{d B}{dt} \approx  4\sqrt{3} \xi E + 2 \sqrt{2} \xi G,
\]
\begin{equation}
i \frac{d E}{dt} \approx 4 \sqrt{3} \xi B  +  2\sqrt{6} \xi G,
\end{equation}
\[
i \frac{d G}{dt} \approx 2 \sqrt{2} \xi B  +  2\sqrt{6} \xi E + 4 \sqrt{3} \xi J,
\]
\[
i \frac{d J}{dt} \approx 4 \sqrt{3} \xi G,
\]
for $D(t)$,
\begin{equation}
i \frac{d D}{dt} \approx 0,
\end{equation}
and for  $C(t)$ and $H(t)$
\[
i \frac{d C}{dt} \approx 2 \sqrt{2} \xi H,
\]
\begin{equation}
i \frac{d H}{dt} \approx 2 \sqrt{2}  \xi C.
\end{equation}
The initial conditions for the state (D1) correspond to $A(0) =a^3$, $B(0) = \sqrt{3} a^2 b$, $C(0) = \sqrt{3} a b^2$, and $D(0)=b^3$. Therefore, the probability amplitudes in the different sectors are separately conserved and so $|A(t)|^2 + |F(t)|^2 + |K(t)|^2 =|a|^6$,  $|B(t)|^2 + |E(t)|^2 + |G(t)|^2  +|J(t)|^2 =3 |a|^4 |b|^2$,  $|D(t)|^2 = |b|^6$, and $|C(t)|^2 + |H(t)|^2 = 3 |a|^2 |b|^4$, which when summed gives the overall probability of unity. Note that if $a=0$, then the three cavity system remains in its initial state viz., $|e,0\rangle^{(1)}|e,0\rangle^{(2)}|e,0\rangle^{(3)}$. However, if $b=0$, viz., the initial state is $|g,2\rangle^{(1)}|g,2\rangle^{(2)}|g,2\rangle^{(3)}$, then the only nonzero amplitudes are $A(t)$, $F(t)$, and $K(t)$ with normalization condition  $|A(t)|^2 + |F(t)|^2 + |K(t)|^2 = 1$.

The general solution of the system of equations (D4)--(D7), with initial condition (D1), are for $A(t)$, $F(t)$, and $K(t)$,
\[
A(t) =\frac{1}{11} a^3 \Big{[} 6 \cos ( 2\sqrt{66} \xi t )+5 \Big{]},
\]
\begin{equation}
F(t) = -\frac{1}{11} a^3  \sqrt{66} i \sin (2 \sqrt{66} \xi t ),
\end{equation}
\[
K(t) = \frac{1}{11} \sqrt{30}  a^3 \Big{[}\cos(2 \sqrt{66} \xi t ) -1 \Big{]}.
\]

For $B(t)$, $E(t)$, $G(t)$, and $J(t)$, one has exact solutions; however, we present their simpler numeric rather than the purely analytic results and so
\[
B(t) =a^2 b\Big{[} 0.4054 e^{-11.2644 i \xi  t} +0.3995 e^{ -3.7306 i \xi t}
\]
\[
+0.0838 e^{8.6745 i \xi t}+ 0.8433 e^{6.3205 i \xi t} \Big{]},
\]
\begin{equation}
E(t) = a^2 b \Big{[} 0.4607 e^{-11.2644 i \xi t} + 0.3401 e^{-3.7306 i \xi t}
\end{equation}
\[
- 0.2040 e^{8.6745 i \xi t} - 0.5968 e^ {6.3205 i \xi t}\Big{]},
\]
\[
G(t) = a^2 b  \Big{[}0.4860 e^ {-11.2644 i \xi t}  -.3061 e^{-3.7306 i \xi t}
\]
\[
+ 0.2427 e^{ 8.6745 i \xi t} -.4227 e ^{6.3205 i \xi t} \Big{]},
\]
\[
J(t)=  a^2 b \Big{[} 0.2989 e^{ -11.2644 i \xi t} -0.5684 e^ {- 3.7306 i \xi t}
\]
\[
- 0.1939 e ^{8.6745 i \xi t} + 0.4633 e^{ 6.3205 i \xi t} \Big{]}.
\]

For $D(t)$ one has that,
\begin{equation}
D(t) = b^3.
\end{equation}

Finally, for $C(t)$ and $H(t)$ one has that,
\[
C(t) =\sqrt{3} a b^2 \cos(2 \sqrt{2} \xi t),
\]
\begin{equation}
H(t) = - \sqrt{3} a b^2 i \sin(2 \sqrt{2} \xi t).
\end{equation}

\section{Asymmetric N=6}

In Appendix D, the case of a symmetric initial state in all three cavities (D1) was considered in the limit of large hopping strength $\xi\gg 1$.  In this Appendix, we consider, in the same large hopping limit, the case where the atom is in the initial unentangled asymmetric state $ |\psi(0)\rangle =|e,2\rangle^{(1)}|g,2\rangle^{(2)}|g,0\rangle^{(3)}$, viz. $D(0)=1$ (see (E1) below). Note that in the limit of large hopping, only photons are transferred between cavities with no ``transfer" of atomic excitations; therefore, the atom in cavity 1 remains in the excited state and so the state of the system at later times is given by
\[
|\psi(t)\rangle = A(t) |e,0\rangle^{(1)}|g,2\rangle^{(2)}|g,2\rangle^{(3)}  \Big{]}
\]
\[
+ |e,0\rangle^{(1)}  \Big{[} B(t)|g,4\rangle^{(2)}|g,0\rangle^{(3)} + C(t)|g,0\rangle^{(2)}|g,4\rangle^{(3)}\Big{]}
\]
\begin{equation}
+|e,2\rangle^{(1)}  \Big{[} D(t)|g,2\rangle^{(2)}|g,0\rangle^{(3)} + E(t)|g,0\rangle^{(2)}|g,2\rangle^{(3)} \Big{]}
\end{equation}
\[
+ F(t) |e,4\rangle^{(1)}|g,0\rangle^{(2)}|g,0\rangle^{(3)}.
\]

The equations that govern the time dependence of the probability amplitudes in (E1) are
\[
i  \frac{d A}{dt} \approx  \sqrt{24}  \xi B + \sqrt{24}  \xi C  + 2 \xi D + 2 \xi E,
\]
\[
i  \frac{d B}{dt} \approx  \sqrt{24}  \xi A + \sqrt{24}  \xi D,
\]
\begin{equation}
i  \frac{d C}{dt} \approx  \sqrt{24}  \xi A + \sqrt{24}  \xi E,
\end{equation}
\[
i  \frac{d D}{dt} \approx   2 \xi A + \sqrt{24}  \xi B + 2 \xi E + \sqrt{24}  \xi F,
\]
\[
i  \frac{d E}{dt} \approx   2 \xi A + \sqrt{24}  \xi C + 2 \xi D + \sqrt{24}  \xi F,
\]
\[
i  \frac{d F}{dt} \approx  \sqrt{24}  \xi D + \sqrt{24}  \xi E.
\]
The state (E1) can be written as the sum of symmetric and antisymmetric states under the exchange $2\leftrightarrow 3$. The probability amplitudes $A(t)$, $B(t) + C(t)$, $D(t) + E(t)$, and $F(t)$ are the symmetric amplitudes with eigenfrequencies $-8\xi$, $-6\xi$, $4\xi$, and $12\xi$.  The amplitudes $B(t)-C(t)$ and $D(t)-E(t)$ are the antisymmetric amplitudes with eigenfrequencies $-6\xi$ and $4\xi$. Accordingly, (E2) can be written as a set of two  and four uncoupled equations for the antisymmetric and symmetric amplitudes, respectively.

The solution of (E2), for the initially unentangled state $|\psi(0)\rangle =|e,2\rangle^{(1)}|g,2\rangle^{(2)}|g,0\rangle^{(3)}$, that is, $D(0)=1$, is
\[
A(t)= E(t) = \frac{1}{15} \Big{[} -2 e^{-4 i\xi t} - 3 e^{6 i \xi t} +  2 e^{8 i \xi t}  + 3 e^{-12 i \xi t} \Big{]},
\]
\begin{equation}
B(t) = F(t) = \frac{\sqrt{6}}{15} \Big{[} e^{-4 i \xi t} -  e^{6 i \xi t} -  e^{8 i \xi t} +  e^{-12 i \xi t} \Big{]},
\end{equation}
\[
C(t) = \frac{\sqrt{6}}{15} \Big{[} -2 e^{-4 i \xi t}  + 2 e^{6 i \xi t} -  e^{8 i \xi t} +  e^{-12 i \xi t} \Big{]},
\]
\[
D(t) = \frac{1}{15} \Big{[} 4 e^{-4 i\xi t} + 6 e^{6 i \xi t} +  2 e^{8 i \xi t}  + 3 e^{-12 i \xi t} \Big{]}.
\]

\begin{newpage}

\bibliography{basename of .bib file}

\end{newpage}

\end{document}